\documentclass[fleqn,usenatbib]{mnras}

\usepackage{mathptmx}

\usepackage[T1]{fontenc}
\usepackage{ae,aecompl}

\usepackage{graphicx}	
\usepackage{amsmath}	
\usepackage{amssymb}	
\usepackage{subfigure}

\newcommand\Msol{M$_{\odot}$}
\newcommand\Lsol{L$_{\odot}$}

\defcitealias{sesar15}{S15}
\defcitealias{bernard14}{B14}

\title[The Ophiuchus stream progenitor]{The Ophiuchus stream progenitor: a new type of globular cluster and its possible Sagittarius connection}

\author[J. M. M. Lane et al.]{James M. M. Lane$^{1,2}$\thanks{E-mail: lane@astro.utoronto.ca},
Julio F. Navarro$^{2}$,
Azadeh Fattahi$^{2,3}$,
Kyle A. Oman$^{2,4}$,
\newauthor
Jo Bovy$^{1}$
\\
$^{1}$ Department of Astronomy \& Astrophysics, University of Toronto, 50 St. George Street, Toronto, ON M5S 3H4, Canada\\
$^{2}$ Department of Physics \& Astronomy, University of Victoria, 3800 Finnerty Road, Victoria, BC V8P 5C2, Canada\\
$^{3}$ Institute for Computational Cosmology, Department of Physics, University of Durham, South Road, Durham DH1 3LE, UK\\
$^{4}$ Kapteyn Astronomical Institute, University of Groningen, Postbus 800, NL-9700 AV Groningen, The Netherlands\\
}

\date{Accepted XXX. Received YYY; in original form ZZZ}

\pubyear{2019}

\begin{document}
\label{firstpage}
\pagerange{\pageref{firstpage}--\pageref{lastpage}}
\maketitle

\begin{abstract}
The Ophiuchus stream is a short arc-like stellar feature of uncertain origin located $\sim 5$ kpc North of the Galactic centre. New proper motions from the second \textit{Gaia} data release reconcile the direction of motion of stream members with the stream arc, resolving a puzzling mismatch reported in earlier work. We use N-body simulations to show that the stream is likely only on its second pericentric passage, and thus was formed recently. The simulations suggest that the entire disrupted progenitor is visible in the observed stream today, and that little further tidal debris lies beyond the ends of the stream. The luminosity, length, width, and velocity dispersion of the stream suggest a globular cluster (GC) progenitor  substantially fainter and of lower surface brightness than estimated in previous work, and unlike any other known globulars in the Galaxy. This result suggests the existence of clusters that would extend the known GC population to fainter and more weakly bound systems than hitherto known. How such a weakly-bound cluster of old stars survived until it was disrupted so recently, however, remains a mystery. Integrating backwards in time, we find that the orbits of Sagittarius and Ophiuchus passed within $\sim 5$ kpc of each other about $\sim 100$ Myrs ago, an interaction that might help resolve this puzzle. \end{abstract}

\begin{keywords}
globular clusters: general -- Galaxy: evolution -- Galaxy: kinematics and dynamics -- Galaxy: structure -- Galaxy: halo -- galaxies: dwarf
\end{keywords}



\section{Introduction}
\label{sec:Introduction}

Recent wide-field imaging campaigns have revealed numerous examples of stellar streams in the halo of the Milky Way (MW) that span a wide range of scales, from the wide remains of the Sagittarius dwarf spheroidal (dSph), which wraps more than once around the sky \citep{majewski03,belokurov06}, to relatively short, thin tails that emerge from globular clusters (GCs) such as Palomar 5 \citep{odenkirchen01}.

Kinematically cold, thin streams are particularly interesting, since they place the tightest contraints on the MW gravitational potential \citep[e.g.][]{bovy16}. Their morphology  may also provide clues to the existence of low-mass dark matter sub-haloes which, although invisible, may induce stream `gaps' through gravitational interaction \citep{ibata02,johnston02,bovy17}.

The Ophiuchus stream is a short, thin overdensity of stars discovered by \citet[][hereafter B14]{bernard14} in the Pan-STARRS1 3$\pi$ survey \citep{kaiser10}. The stream is located at $(l,b) \approx (4\fdg5,+32\degr)$, and subtends $\sim 2.5$~degrees in length and $\sim 7$~arcmin in width. The main sequence is clearly identifiable in deep color-magnitude diagrams of the region, although the sparsely populated red giant branch is barely discernible amid the stellar foreground/background. \citetalias{bernard14} found that the stream's colour-magnitude profile was well approximated by the isochrone of an old metal-poor globular cluster (namely NGC~5904) at a distance of $\sim 9.5$ kpc from the Sun, suggesting a tidally disrupted globular cluster.

\citet[][hereafter S15]{sesar15} obtained spectra for $\sim 170$ potential stream stars, out of which 14 were identified as  stream members based on their radial velocities. These authors concurred that  the stream likely originated from a metal-poor globular cluster with an age and $[\text{Fe/H}]$ of about and 11.7~Gyr and -1.95, respectively. They concluded that the stream, which lies almost directly north of the Galactic centre, is highly forshortened in projection, with a true length of about 1.5~kpc. Using radial velocities and proper motions, they integrated an orbit for the stream in a Milky Way-like potential and inferred that it must have disrupted around 240~Myr ago.

Such recent disruption is difficult to reconcile with the old ages of Ophiuchus' constituent stars. No bound core has been identified within its extent or near its orbital path, suggesting that the stream is highly evolved and has completed a number of orbits around  the Galaxy. In contrast, the short deprojected length of the stream suggests the opposite; i.e., that the stream is dynamically young, and has completed very few orbits. This paradox has motivated a number of possible scenarios.

One is that the stream has been shortened by the gravitational influence of the Galactic bar. Indeed, \citet{sesar16} identified 4 blue horizontal branch stars projected near the end of the stream that have radial velocities distinct from the stream, but still unusual for halo stars at that location ($v_{\rm los} > 230$ km~s$^{-1}$). These authors interpreted those stars as stream members that may have `fanned out' through non-linear interactions with the bar. Stream `fanning' may be enough to disperse the stream ends below detectability, causing the stream to appear shorter than it truly is. \citet{price-whelan16} reached similar conclusions after exploring the effects of bar-induced chaotic orbits on the properties of a stream like Ophiuchus. Finally, \citet{hattori16} argued that the bar may have a `shepherding' effect on the Ophiuchus stream, allowing it to remain at a fixed length for 1~Gyr or more, which is many times longer than the disruption time predicted by S15.

While these studies suggest that the Galactic bar may have played an important role on the evolution of the Ophiuchus stream, their results are highly sensitive to the mass of the bar and its exact pattern speed, as well as to the dynamical age and previous evolution of the stream, none of which are known well enough to reach definitive and reliable conclusions.

\begin{figure}
	\includegraphics[width=\columnwidth]{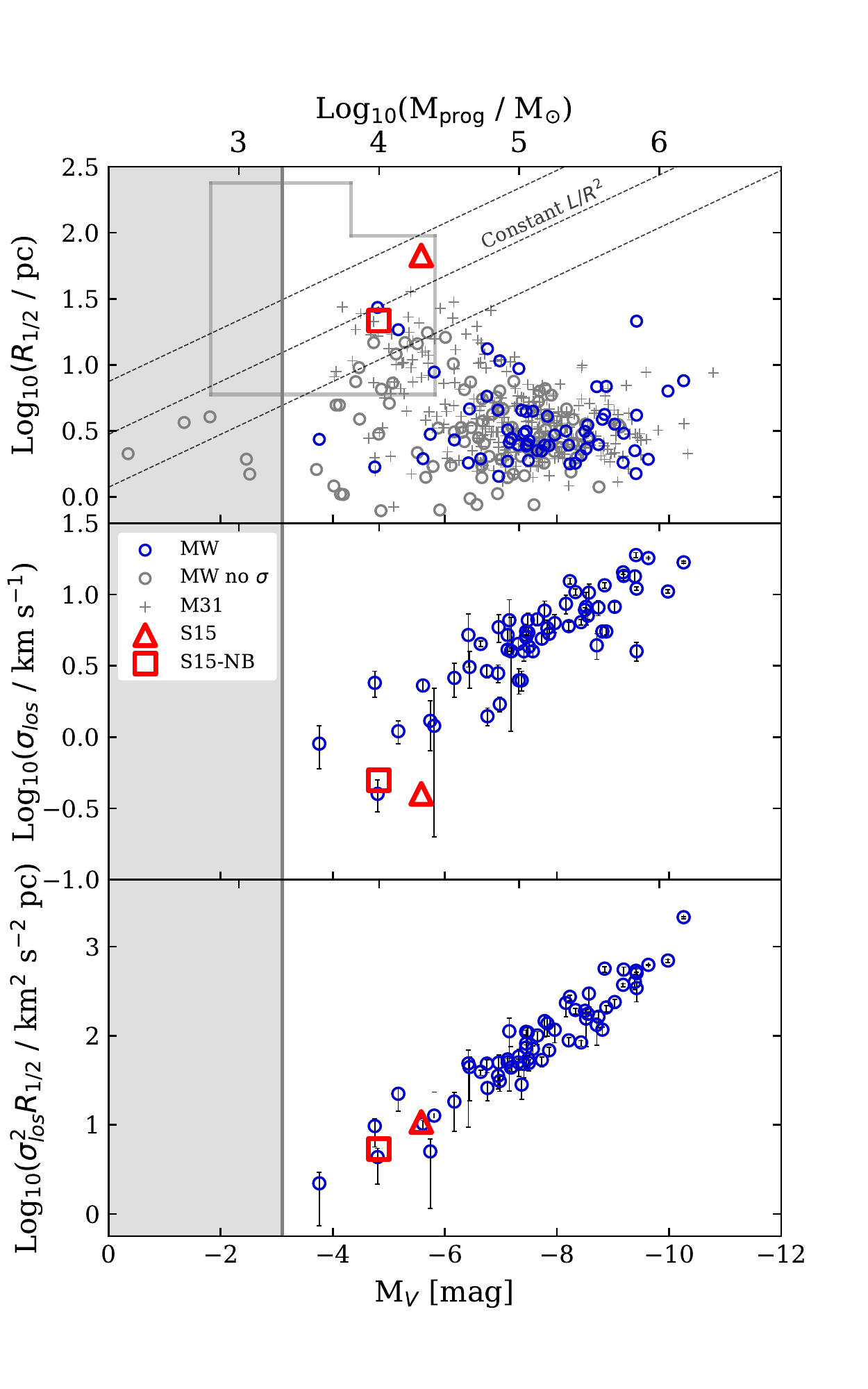}
    \caption{The globular clusters of the Milky Way \citep[circles, from][]{harris96}, M31 \citep[crosses, from][]{huxor14,peacock10}, and the S15 and S15-NB progenitors shown in three parameter spaces as functions of their absolute magnitude. Clusters with blue symbols have line-of-sight velocity dispersion measurements. \textit{Top:} projected half-light radius as a function of the absolute magnitude. The grey dashed lines show constant surface brightness. The grey outline shows the range of progenitor properties studied in this work. \textit{Middle:} Velocity dispersion as a function of the absolute magnitude for those clusters which have velocity dispersion measurements. \textit{Bottom:} Dynamical mass as a function of the absolute magnitude for those clusters which have velocity dispersion measurements. The cluster mass shown on the top axis is calculated assuming a mass-to-light ratio of 1.45. In all three panels the grey shaded box shows the range of magnitudes excluded by the observed luminosity of the stream.}
    \label{fig:GlobularClusters}
\end{figure}

A simpler alternative is that the progenitor was originally so weakly bound that it completely disrupted in just a few orbits, leaving behind a short tidal tail and no bound core. This is indeed the scenario explored by S15, who estimated for the progenitor a stellar mass of $\sim 2\times 10^{4}$ \Msol, and a velocity dispersion of $\sim 0.4$~km~s$^{-1}$. These properties imply a rather large size, unusual for a typical GC. Another difficulty is that a system so weakly bound cannot have orbited the Galaxy in its present orbit more than a few times, raising questions about its origin. Presumably the progenitor formed in a very different orbit and has only recently, perhaps as a result of interactions with a Galactic satellite, reached its present-day orbit.

The work presented here examines these issues further by carefully analyzing the tidal remnants of a large number of possible progenitors, spanning a large range in GC stellar mass and size/velocity dispersion. Detailed comparison with observations allows us to revise earlier constraints on these parameters, suggesting that the most likely progenitor GC was even more unusual in its properties, deepening the mystery of its origin. Although we do not consider the effects of the Galactic bar in this work, we do explore the possibility that Ophiuchus may have interacted in the recent past with the Sagittarius dwarf, offering a possible clue to the resolution of this puzzle in future work.

The paper is arranged as follows: in Section~\ref{sec:Methods} we describe our simulations and models for both the cluster and the Milky Way potential. Section~\ref{sec:ObservingSimulatedStreams} describes the stream analysis procedure, while Section~\ref{sec:Results} explains how the properties of simulated streams are derived. Section~\ref{sec:SgrOPhiuchusInteraction}, in particular, explores the possibility that Ophiuchus has interacted with the Sagittarius dwarf spheroidal (dSph). We summarize our findings in Section~\ref{sec:SummaryAndDiscussion}.

\begin{figure}
	\includegraphics[width=\columnwidth]{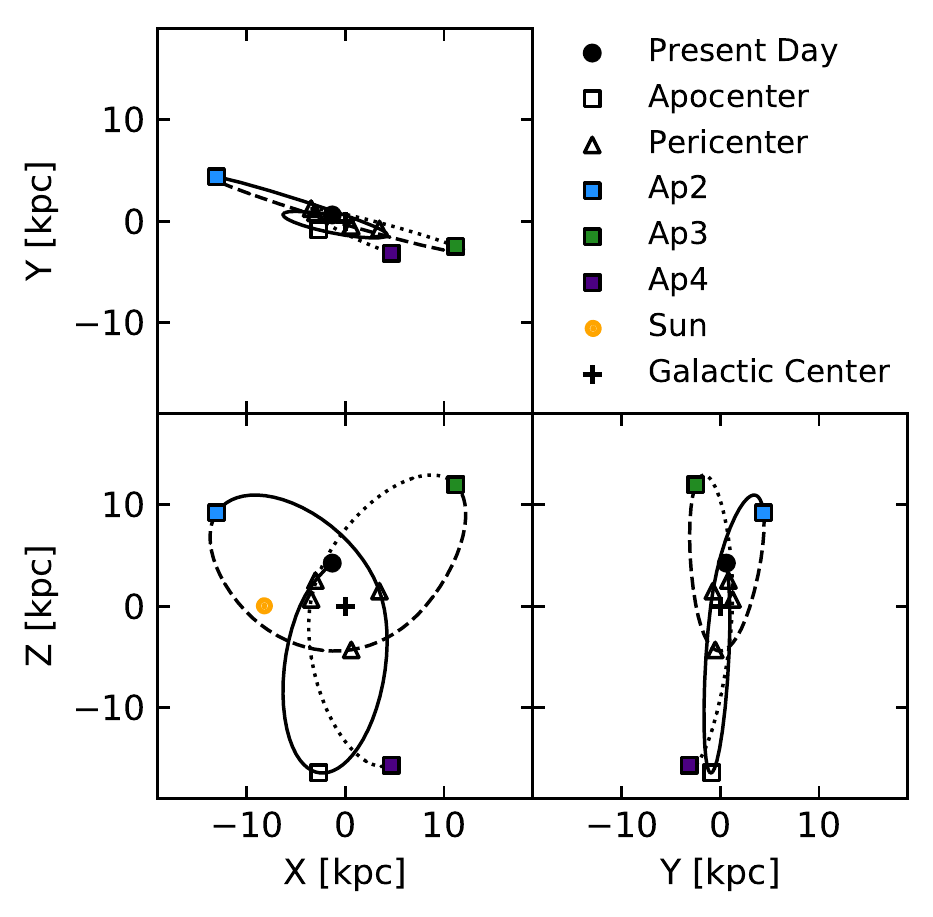}
    \caption{Orbit of the Ophiuchus stream over the last $837$~Myr in galactocentric Cartesian coordinates. Squares and triangles mark apocentric and pericentric passages, respectively. The coloured squares are the apocentres where we begin simulations. The solid, dashed, and dotted lines cumulatively show the orbit starting from the second, third and fourth most recent apocentric passages. The black circle marks the present-day location of the stream. The black cross and orange dot mark the positions of the Sun at $(-8.3,0,0)$, and the Galactic centre at $(0,0,0)$, respectively.}
    \label{fig:Orbit}
\end{figure}

\section{Numerical Simulations}
\label{sec:Methods}

\subsection{GC models}
\label{subsec:GCModel}

We model each globular cluster progenitor as a $10^{4}$ particle realization of a \citet{plummer11} model with density profile, $\rho(r)=\rho_{\rm P}/(1+r^2/r^2_{\rm P})^{5/2}$, where $\rho_{\rm P}=3M/(4\pi r_{\rm P}^3)$. The N-body initial conditions are realized using the Zeno toolkit\footnote{\url{https://github.com/joshuabarnes/zeno}}. We assume that the GC contains only stars, which implies that its physical properties are set by the mass, $M$, and scale radius, $r_{\rm P}$, of our model, from which the velocity dispersion follows. We set the gravitational softening to $0.15$ times the Plummer scale radius, and allow the N-body system to relax in isolation for many cluster crossing times before evolving it in the Galactic potential.

\subsubsection{The S15 progenitor models}

We begin by considering the progenitor models presented in \citetalias{sesar15}. These authors consider two different models, whose properties are listed in Table~\ref{table:progenitors} and are shown, by the red square and triangle, in Figure~\ref{fig:GlobularClusters}. Masses in Figure~\ref{fig:GlobularClusters} refer to the total stellar mass of the cluster ($M_V$ is the absolute magnitude assuming a mass-to-light ratio of $1.45$~\Msol~\Lsol$^{-1}$), R$_{1/2}$ is the 2D projected half-mass radius (calculated as 3/4 times the 3D half-mass radius), and $\sigma_{\rm los}$ is the line-of-sight velocity dispersion.

\begin{table}
	\centering
	\caption{Properties of the progenitor globular clusters from \citetalias{sesar15}.}
	\label{table:progenitors}
	\begin{tabular}{cccc}
		\hline
		Progenitor Name	&	Mass	 (\Msol)		&	r$_{1/2}$ (pc)	&	$\sigma_{\rm vlos}$ (km~s$^{-1}$)	\\
		\hline
		S15				&	$2\times10^{4}$		&	90			&	0.40			\\
		S15--NB			&	$1\times10^{4}$		&	29			&	0.50		\\
		\hline
	\end{tabular}
\end{table}

The properties of the progenitor labelled `S15' are taken from table~1 of \citetalias{sesar15}. The half-mass radius is derived following equation (2) of \cite{wolf10} for the relationship between mass, radius and velocity dispersion in spherical, dispersion-supported systems. Model S15-NB is a King profile of mass of $10^{4}$~\Msol, tidal radius of 94~pc, and ratio of central potential to velocity dispersion squared of 2.0. These properties imply a concentration parameter of 0.5 \citep[see figure~4.9 in][]{binney08} and therefore a half-mass radius of approximately 29~pc. Using the mass-radius-velocity dispersion relations of \citet{wolf10} this implies a velocity dispersion of 0.5~km~s$^{-1}$.

\subsection{GC model grid}

In addition to S15 and S15-NB, we explore a grid of GC models in the space of total mass and half-mass radius. More specifically, we consider Plummer models with half-mass radii between $10$~pc and $100$~pc, and masses between $8 \times 10^{2}$~\Msol\ and $2 \times 10^4$~\Msol.  We sample this range of parameters, shown in Figure~\ref{fig:GlobularClusters}, in 0.2 dex intervals. In addition, for clusters less massive than $5\times10^{3}$~\Msol \,  we also examine radii up to 250~pc,  for a total of $58$ candidate progenitors.

The lower mass boundary is motivated by the total luminosity of the stream, which according to \citetalias{bernard14}, is $\sim 1.4\pm0.6\times10^{3}$~L$_{\odot}$. At a mass of $2 \times 10^{4}$~\Msol\ the selected range of half-mass radii correspond to a range of line-of-sight velocity dispersions between 0.37 and 1.2 km~s$^{-1}$, and at $2 \times 10^{3}$~\Msol\ the range of dispersions is between 0.11 and 0.37 km~s$^{-1}$.  This range comfortably spans the 68~per~cent confidence interval derived by \citetalias{sesar15} for the velocity dispersion of the progenitor.

\begin{figure}
	\includegraphics[width=\columnwidth]{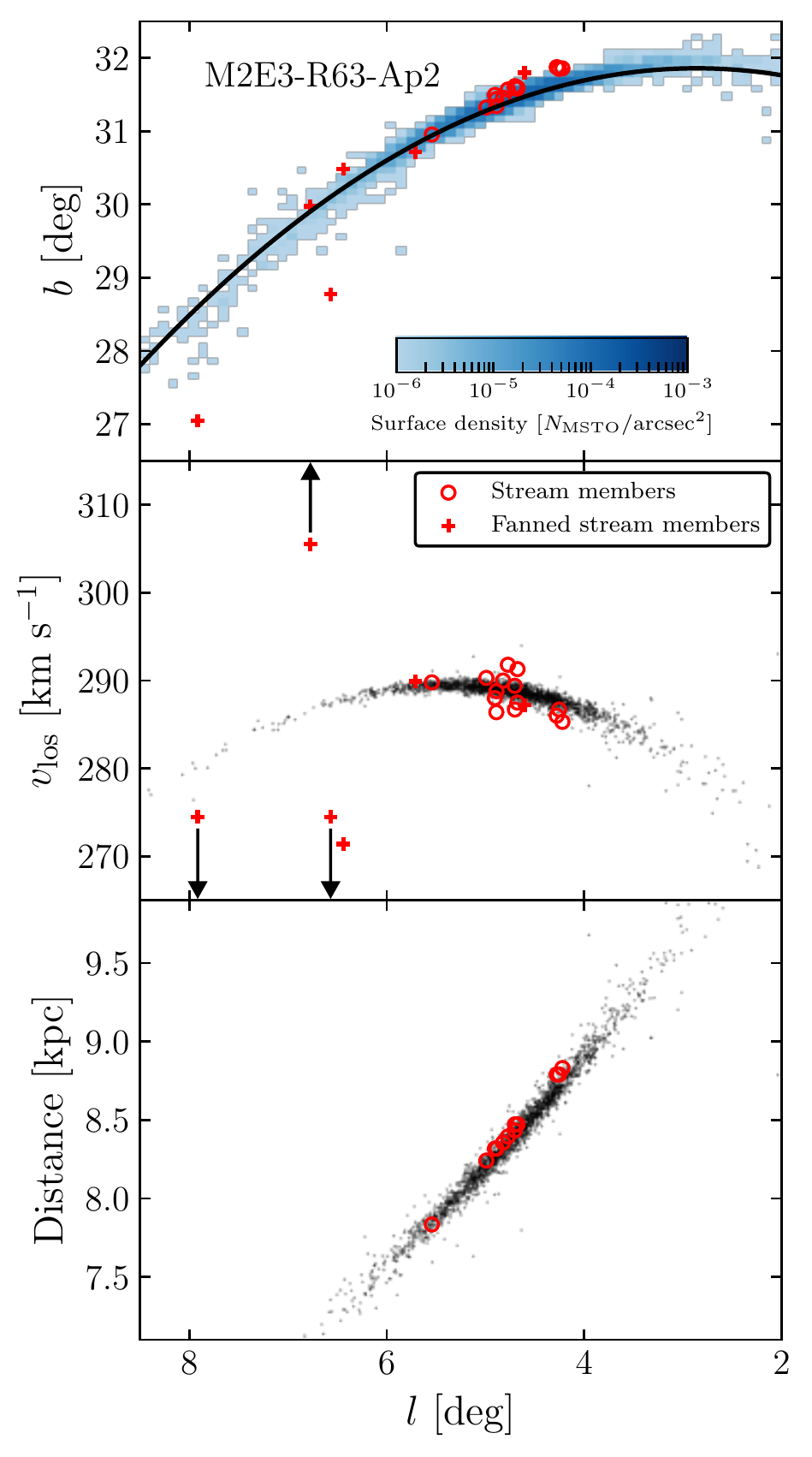}
    \caption{Kinematics of the M2E3-R63-Ap2 progenitor stream as a function of Galactic longitude. The top, middle and bottom panels show the Galactic latitude, heliocentric radial velocity, and distance. The colour scale in the top panel shows number surface density of N-body particles expressed as \textit{MSTO} stars, while in the bottom two panels the particles are individually shown as black dots. The red circles are the confirmed stream members from \citetalias{sesar15}. The thick black line in the top panel is the best-fitting quadratic to the stream extent. The red crosses show the candidate fanned stream members from \citet{sesar16}, which do not have measured distances. Arrows mark fanned stream candidates that lie outside of the plotting window. Our simulated streams match the observations well in all of the observed coordinates.
    } 
    \label{fig:Banana}
\end{figure}

\subsection{Galactic potential and progenitor orbits}
\label{subsec:GalModel}

To model of the Galactic potential we follow \citetalias{sesar15} and use the 3-component Milky Way potential \texttt{MWPotential2014} from the galactic dynamics package \texttt{galpy}\footnote{\url{https://github.com/jobovy/galpy}} \citep{bovy15}. This potential consists of a \citet{miyamoto75} disc, an exponentially truncated power law density profile for the bulge, and an NFW halo \citep{navarro97}. For a full list of the physical parameters that describe this model we refer the reader to section~3.5 and table~1 of \citet{bovy15}.

S15 report that the stream traces an orbit in this potential, with consistent radial velocities. They also estimated proper motions using 2MASS and archival photographic plate observations and reported that, for the inferred distance of the stream, the resulting 3D velocities were misaligned with the stream, suggesting an inconsistency between the stream and the Galactic model. However, accurate proper motions for the stream stars have recently become available from the \textit{Gaia} second data release \citep[DR2][]{gaia18a}. We have obtained proper motions for the 14 stream members from the \textit{Gaia} DR2 archive\footnote{\url{https://gea.esac.esa.int/archive/}} and found them to be consistent with the orbit of \citetalias{sesar15}, neatly resolving this tension. For more information about stream member kinematics from \textit{Gaia} DR2 see Appendix~\ref{ap:Gaia}. The orbital parameter values are summarized in Table~\ref{table:kinematics}; we refer the reader to S15 for a full discussion of their derivation and the associated uncertainties.

\begin{table}
	\centering
	\caption{Present-day kinematics of the Ophiuchus stream from \citetalias{sesar15}.}
	\label{table:kinematics}
	\begin{tabular}{cc}
		\hline
		Parameter		&	Value			\\
		\hline
		$l$			&	5$\degr$			\\
		$b$			&	31.37$\degr$		\\
		$d_{\odot}$			&	8.2 kpc			\\
		$v_{\rm los}$		&	289.1 km~s$^{-1}$	\\
		$\mu_{l}$		&	-7.7 mas~yr$^{-1}$	\\
		$\mu_{b}$		&	1.4 mas~yr$^{-1}$	\\
		\hline
	\end{tabular}
\end{table}

\begin{figure*}
	\includegraphics[width=\textwidth]{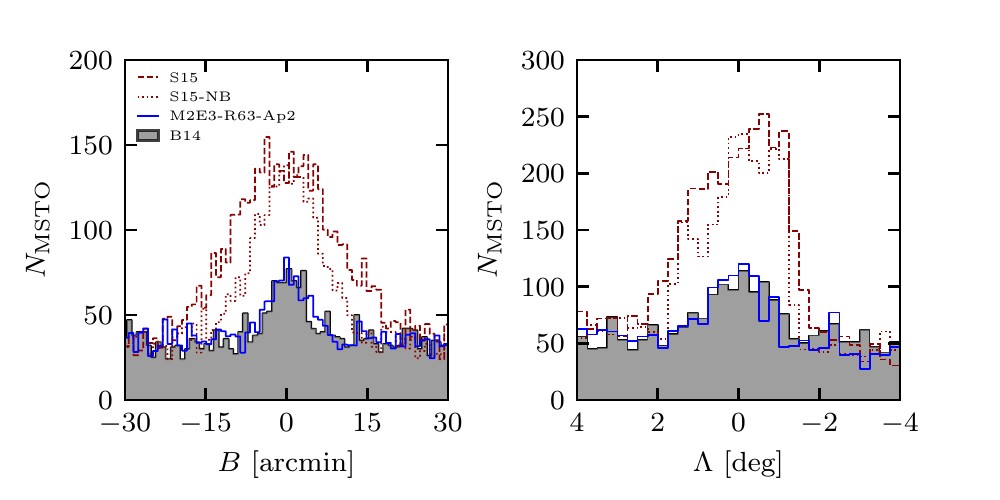}
   	 \caption{Histograms of MSTO star counts along the length and width of the simulated streams. The dark red dashed and dotted histograms show the S15 and S15-NB models, respectively. The black shaded histogram shows the observations from \citetalias{bernard14}. For comparison the stream M2E3-R63-Ap2 is shown in blue, which is clearly a better match to the data than the S15 or S15-NB models.}
   	 \label{fig:Dimensions}
\end{figure*}

\subsection{Simulations}
\label{subsec:Simulations}

To simulate the disruption of the progenitors of the  Ophiuchus stream we use the {\sc Gadget-2} code \citep{springel05}, after including a static \texttt{MWPotential2014} potential. The initial conditions for the simulations were derived using the present-day orbit discussed above, after evolving it backwards in time for 4 full radial periods. (The radial period of the orbit is 240~Myr.)

Figure~\ref{fig:Orbit} shows the orbital path over the last $837$~Myr. We ran each of our GC models three times, starting at the second, third, or fourth most recent apocentric passage, denoted as Ap2, Ap3 and Ap4, respectively. The locations of their starting points are marked with the coloured squares in Figure~\ref{fig:Orbit}. The duration of the simulations are $t=361$, $601$, and $837$~Myr for Ap2, Ap3 and Ap4, respectively. These integrations are short enough that our use of a static Galactic potential is a reasonable approximation to the Milky Way potential over time. Each simulation is halted when the stream reaches its present day position for comparison with the observed stream. Throughout the remainder of the paper our naming convention is such that, for example,  M2E3-R63-Ap2 refers to a progenitor mass of $2 \times 10^{3}$~\Msol, half mass radius of 63~pc, evolved from the second most recent apocentre.

\section{Simulated vs Observed Ophiuchus Stream}
\label{sec:ObservingSimulatedStreams}

We mock-observe our simulated streams by converting them to Galactic coordinates and observing them from the Sun's location. Particles in the simulation are used to render actual stars using a Chabrier IMF and a sampling procedure described in detail in Appendix~\ref{ap:ParticleMassConversion}. This procedure allows us to associate total stellar mass at some sky location with a direct observable, such as the total number of main sequence turnoff (MSTO) stars. In Appendix~\ref{ap:MassFunctionFlattening} we assess the potential impact of the flattening of the stellar mass function due to the impact of tides and the internal dynamical evolution of the cluster.

Figure~\ref{fig:Banana} illustrates the mock observation procedure for one particular stream, M2E3-R63-Ap2.  This figure shows, as a function of Galactic longitude, $l$, the Galactic latitude, $b$, the heliocentric line-of-sight velocity $V_{\rm los}$, and the heliocentric distance of stream stars. Confirmed stream members from \citetalias{sesar15} are shown as red circles.

It is clear that our simulated streams match the overall morphology of the observed stream quite well.  The radial velocities of member stars appear to have greater scatter than the simulated stream, but this is due mainly to observational uncertainties, which are of order $\sim 2$~km~s$^{-1}$ rms. The red crosses in this figure show, for completeness, the `fanned' stream candidates from \citet{sesar16} (three of which have velocities outside the plot limits). We do not expect our models to match the kinematics of these stars.

\subsection{Stream reference frame}

Each of the simulated stream profiles in projection may be approximated by a quadratic polynomial, as shown by the solid line in the top panel of Figure~\ref{fig:Banana}. When fitting the polynomial to the ensemble of N-body particles we weight the fit by the inverse of the projected surface density. Once the polynomial is fit we can rectify the stream to a reference system where parameters like the length and width of the stream can be meaningfully measured and compared with observations. In this  new coordinate system the `latitude' $B$ measures the minimum distance from each star to the fit and the longitude coordinate $\Lambda$ measures the arc length along the quadratic polynomial from a reference position, chosen as the median Galactic longitude of all stream particles. 

Using these new coordinates, and the conversion between N-body particle mass and MSTO stars detailed in Appendix~\ref{ap:ParticleMassConversion}, we calculate below the length and width of our simulated streams following the approach of \citetalias{bernard14} (see their figure~3).

\subsection{Width, length, luminosity, and velocity dispersion estimates}

\subsubsection{Simulated streams}
\label{subsec:SimulatedStreamProperties}

The stream length and width are estimated from histograms of the number of MSTO stars along both the $\Lambda$ and $B$ directions (Figure~\ref{fig:Dimensions}). For the  histogram, all particles between $-1\degr < \Lambda < 1\degr$ are used. Similarly, for $\Lambda$ all particles between $-6\,{\rm arcmin} < B < 6\,{\rm arcmin}$ are used. The gray histogram in each plot is observational data from \citetalias{bernard14}.

A stellar background has been added to each of the simulated streams in order to mimic foreground and background stars in the observations of \citetalias{bernard14}. The purpose of the background is to ensure that our determination of stream parameters is as faithful to those of \citetalias{bernard14} as possible. The background noise is assumed Gaussian with a mean of $N$ and standard deviation of $\sqrt{N}$, where $N$ is estimated from figure~3 in \citetalias{bernard14}. For the figure showing $B$, we use a mean of $N=35$ and for $\Lambda$ we use a mean that decreases linearly from $N=60$ at $\Lambda=5$ to $N=40$ at $\Lambda=-5$ to account for the latitude dependence of foreground stars.

To determine the width of the stream we follow \citetalias{bernard14} and fit a Gaussian to the $B$ histogram using a least-squares method, and take the FWHM (approximately 2.355 times the standard deviation). The length is estimated by starting at the peak of the $\Lambda$ histogram and moving towards both higher and lower values of $\Lambda$ until a bin with a value below the local noise is reached (without considering the background) on each side of the peak. The length is taken as the difference between these two stream--noise limits.

The number of MSTO stars in the stream, $S_{\text{MSTO}}$, is estimated by summing both the $B$ and $\Lambda$ histograms between the stream--noise limits (also calculated for the $B$ histogram but not related to the reported width), after correcting for the expected number of background stars. In practice, the correction involves drawing repeated samples of the foreground and background stars, and averaging the final results. The uncertainty in the resulting mean $S_{\text{MSTO}}$ is much smaller than the observational uncertainty, ensuring that any difference between observed and simulated stream parameters is not due to the artificial background.

The line-of-sight velocity dispersion is determined for each simulated stream by measuring the radial velocity dispersion of the particles in individual $6\,{\rm arcmin} \times 6\,{\rm arcmin}$ bins projected on the sky (the same bins shown in the top panel of Figure~\ref{fig:Banana}). These individual measurements are then weighted by the particle surface density in the bin and averaged to produce a velocity gradient-independent measurement of the line-of-sight velocity dispersion for the whole stream. This is similar to the manner in which \citetalias{sesar15} measured the intrinsic velocity dispersion of the observed stream.

We note that the only uncertainties involved in our analysis arise from the uncertainty in the fit to the $B$ histogram and the determination of the velocity dispersion. The uncertainty in the width arises from the least-squares fit, and for most streams is of order $1~{\rm arcmin}$. The uncertainty in the velocity dispersion is the standard deviation of the individual velocity dispersion samples, and ranges from less than 0.1~km~s$^{-1}$ for lower mass progenitors to about 0.5~km~s$^{-1}$ for higher mass progenitors. By design of the background subtraction scheme the length measurements carry no uncertainties and mean $S_{MSTO}$ measurements carry uncertainties which are less than 10 per cent of their observational counterparts.

\subsubsection{Observed stream}

The observational value of $S_{\text{MSTO}}$ is estimated using a procedure similar to that described above, using figures~3 and 4 of \citetalias{bernard14}. In practice, we add up stars in their figure~3a between $-10\arcmin \leq B \leq +10\arcmin$, and then subtract a constant background of 35 stars per bin. We also add up stars from their figure~3b between $-1\degr \leq \Lambda \leq 1.25\degr$ and subtract a noise profile that varies linearly from 60 at $\Lambda=5\degr$ to 40 at $\Lambda=-5\degr$.  We then average  the two values to obtain $S_{\text{MSTO}}=389\pm57$ for the observed stream. 

For the observed length and width we adopt $2.5\degr$ and $7.0\pm0.8\,{\rm arcmin}$ (Gaussian FWHM), respectively, as reported by \citetalias{bernard14}. We adopt an uncertainty in length of $0.25\degr$ which corresponds to half the width of one bin in figure~3b of \citetalias{bernard14}.

Finally, we use the measured value of $\sigma_{\rm vlos}$ from \citetalias{sesar15}, which is $0.4^{+0.5}_{-0.4}$; the uncertainty is the central 68~per~cent confidence interval of the posterior probability distribution. We note that this is not the inferred velocity dispersion of the stream progenitor, but rather the intrinsic velocity dispersion of the stream, comparable to the measurement performed on the simulated streams as described above.

\section{Results}
\label{sec:Results}

\subsection{Stream progenitors}
\label{sec:ComparingProgenitors}

\begin{figure*}
    \centering
	\subfigure{\includegraphics[width=3.4in]{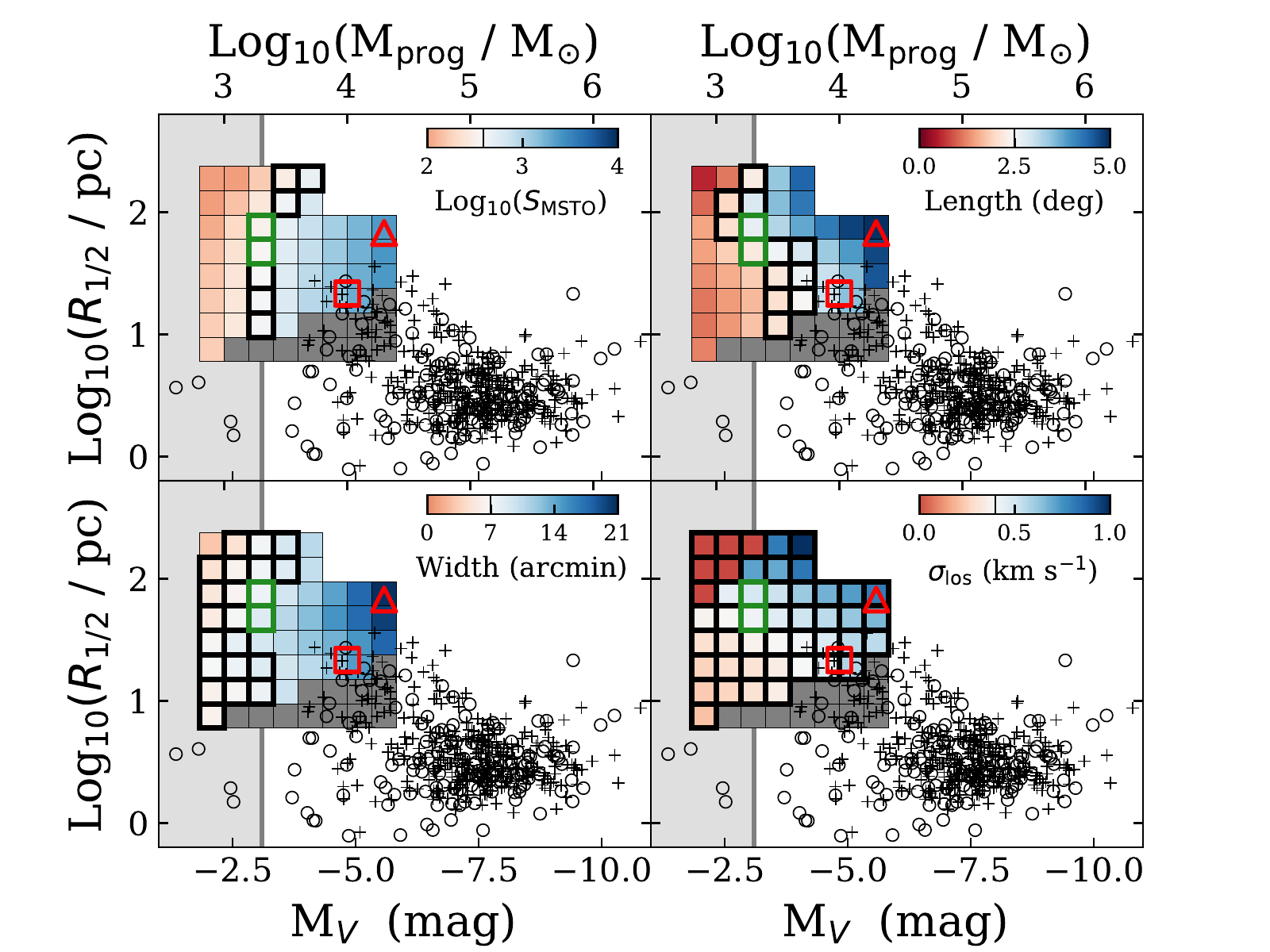}}
	\subfigure{\includegraphics[width=3.4in]{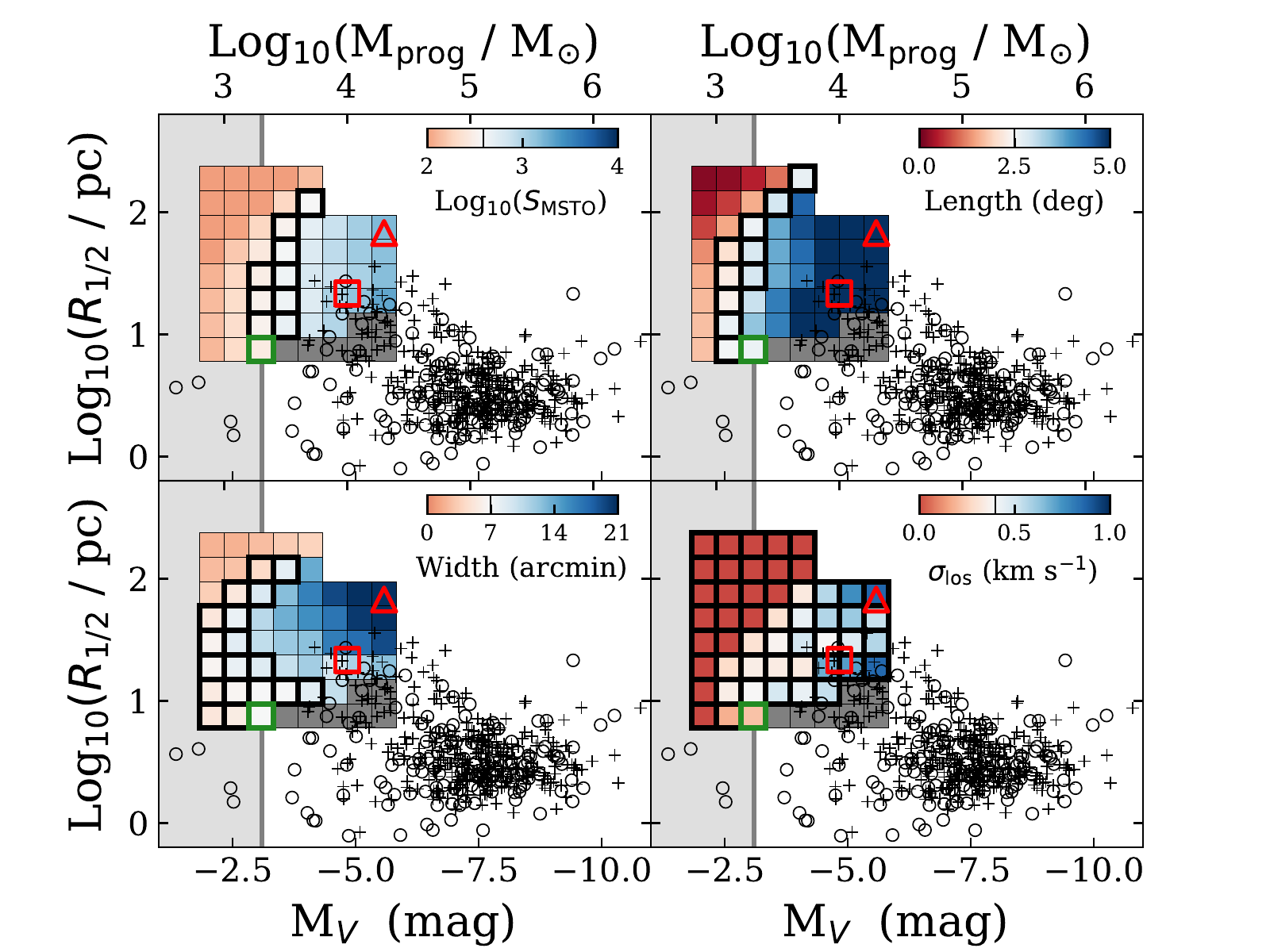}}
   	 \caption{Comparing simulated streams to observations. The values of each derived parameter for our simulated streams for Ap2 (left) and Ap3 (right) models are shown overlaid on a plot of globular cluster total magnitude against logarithmic projected half-light radius. The colouring of the grid in each panel shows one of the derived parameters: logarithmic number of MSTO stars, length, width, and line-of-sight velocity dispersion. The boxes forming the grids are centred on the value of the total magnitude and 2-D half-mass radius of that progenitor. Dark grey cells represent progenitors that can be excluded on morphological grounds. Bolded cells are those in which the parameter value matches the observed value within $2-\sigma$. Green-bordered cells are those in which all four parameters match within $2-\sigma$. The top axis shows progenitor mass assuming a mass-to-light ratio of $1.45$. The black circles and crosses are Milky Way \citep{harris96} and M31 \citep{huxor14,peacock10} globular clusters, respectively. The S15 and S15-NB progenitors are marked using red triangles and squares, respectively. This demonstrates that progenitors best-matched to observations have masses of $2\times10^{3}$~M$_{\odot}$, but a range of potential sizes. }
   	 \label{fig:ParamVals}
\end{figure*}

We assess the viability of different stream progenitors by comparing the integrated number of MSTO stars, $S_{\text{MSTO}}$, the length, width, and the line-of-sight velocity dispersion, $\sigma_{\rm vlos}$ of simulated streams with those of Ophiuchus. Our main results are summarized in Figure~\ref{fig:ParamVals}, where we report how well each of the GC candidates in our model grid is able to match the observed properties of the stream. The comparison is made at second (Ap2, left panel) or third (Ap3, right panel) pericentric passage.  We do not discuss Ap4 models as we find that none of the Ap4 streams provide a convincing match to the Ophiuchus stream. 

The coloured grids in each of the four panels show each one of the measured properties of the resulting stream. The colour bar indicates the value of the parameter for the stream generated by each progenitor, where white has been set to the observed stream parameters. Red or blue thus indicate deviations from observations where the parameter is smaller or larger than observed, respectively. Dark grey indicate progenitors whose streams can be excluded because of obvious morphological considerations, such as cases where the progenitor has not disrupted, or an obvious bound core remains.

To make a quantitative statement about how well our simulated progenitors match the observed stream we need to consider the uncertainties in the measured properties of both simulated and observed streams. We generate a combined uncertainty, defined as the combination in quadrature of both the uncertainties which arise from our analysis (computed as described in Section~\ref{subsec:SimulatedStreamProperties}), and observational uncertainties taken from the literature. For $S_{\text{MSTO}}$ we make the approximation that $\sigma_{\rm{Log}_{10}(S_{\text{MSTO}})} \approx \sigma_{S_{\text{MSTO}}} / (S_{\text{MSTO}} \ln10)$.

In order to visualize our results we highlight in bold progenitors in Figure~\ref{fig:ParamVals} for which the measurement of the respective parameter differs from the observed value by less than two combined standard deviations. The progenitors for which all four parameters match observations in this manner are outlined in green instead.

Figure~\ref{fig:ParamVals} shows that the most discriminating parameters are the length of the stream and $S_{\text{MSTO}}$, with the width also excluding mainly high-mass progenitors. The line of sight velocity dispersion, on the other hand, is a weak discriminant between progenitors, mainly because the observational uncertainties are larger for these progenitors, which have few MSTO stars, therefore inflating the standard deviation.

There are three progenitors which match all four measured parameters within the uncertainties. Two of these are very similar Ap2 progenitors, with masses of order $2\times10^{3}$~\Msol\ and half-mass radii between $\sim 60$ and $100$~pc. The third is an Ap3 model of similar mass but with half-mass radius of $\sim 10$~pc. These progenitors have total luminosity consistent with the luminosity of the stream reported by \citetalias{bernard14}, implying that most of the progenitor is visible in the stream. In contrast, both the S15 and S15-NB models do not match well any of the observed parameters, with the exception of the line-of-sight velocity dispersion.

In Appendix~\ref{ap:MassFunctionFlattening} we demonstrate that if the Ophiuchus progenitor has undergone any mass function flattening, whether due to internal dynamical evolution or external tidal effects, the result will be that we overestimate the inferred mass of the progenitor. We can therefore confidently say that our progenitor mass determination of $2\times10^{3}$~M$_{\odot}$ represents an upper bound. We do not find that progenitor size estimates are affected by mass function flattening, with Ap2 models continuing to favour half-mass radii between $60$ and $100$~pc, and Ap3 models favouring even smaller radii.

Note that there are no known GCs in the Local Group as faint and as weakly-bound as the progenitors that our analysis favours. The only known clusters with similar stellar mass/luminosity have half-mass radii about an order of magnitude smaller than expected for the Ophiuchus progenitor. This is an intriguing finding, as it suggests that the GC population may span a larger range of radii and surface brightness than hitherto known. A progenitor like the one favoured by our modeling would be rather difficult to find, given its vanishingly small surface brightness, but we see no a priori reason to exclude their presence in the Galactic halo, even in large numbers. Taken at face value, our results suggest that our understanding of the faint GC population may be rather incomplete.

\begin{figure}
	\includegraphics[width=\columnwidth]{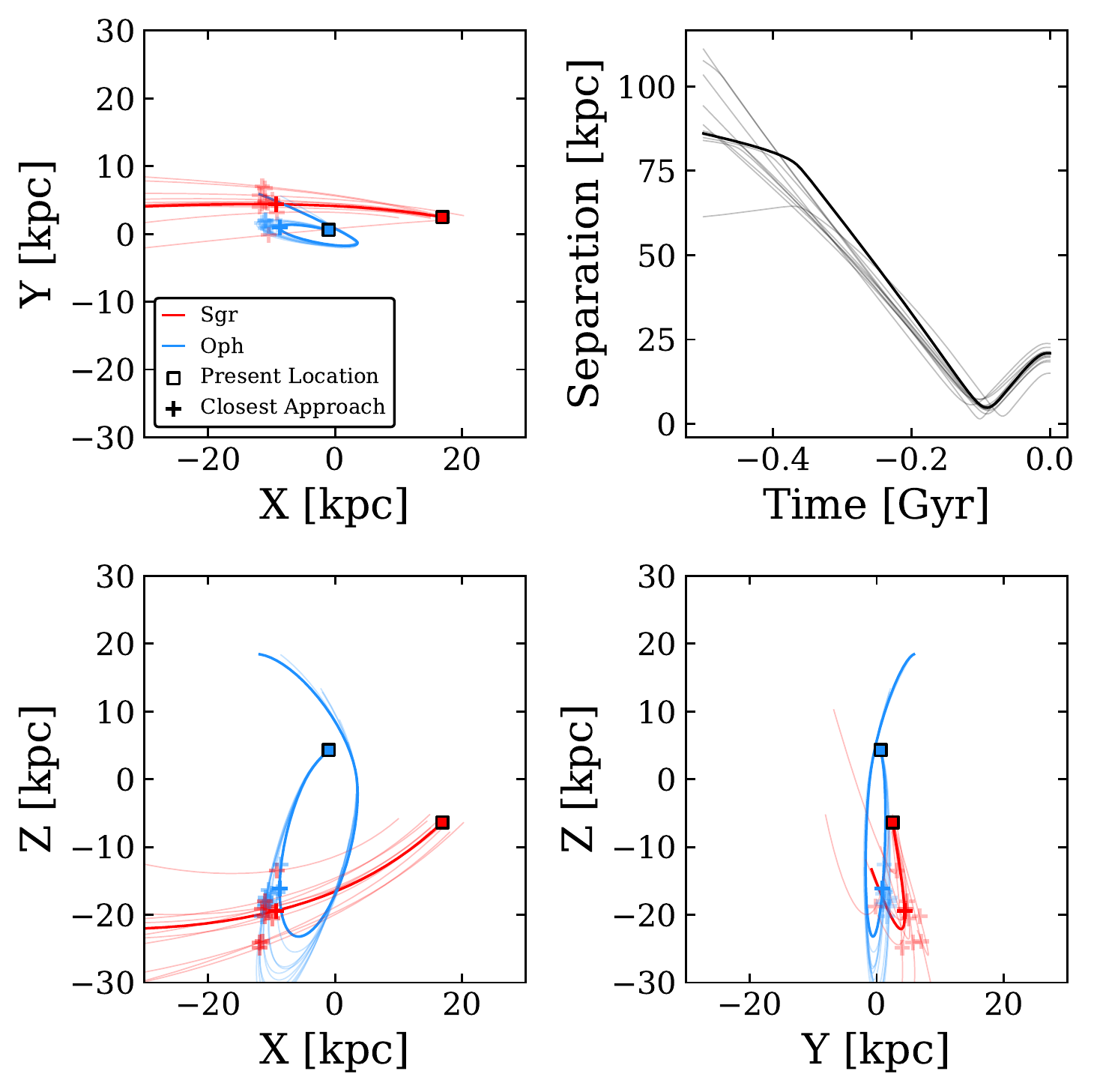}
        \caption{Kinematics of 25 pairs of past orbits of the Ophiuchus progenitor and the Sagittarius dSph (Sgr). The top left and bottom two panels show three orthogonal projections. The Sun is at $(-8.3,0,0)$, and the Galactic centre at $(0,0,0)$, respectively, in these projections. The top right panel shows the separation between the two bodies as a function of time. The bolded orbits are those corresponding to the unsampled (i.e. not sampled from the error distribution) kinematics of Sgr and Ophiuchus. These integrations suggest a close passage between Sgr and Ophiuchus about $100$~Myrs ago, at about the time of the last Ophiuchus apocentric passage.}
    \label{fig:Sgr}
\end{figure}

\subsection{A connection to Sagittarius?}
\label{sec:SgrOPhiuchusInteraction}

The above analysis demonstrates that a very low mass, weakly-bound GC is a viable progenitor for the Ophiuchus stream, but it does not address the question of its origin. Given the age of its stars and the short time it takes to disrupt, it is clear that the Ophiuchus progenitor could not have formed in its present orbit.  One possibility is that the Ophiuchus progenitor cluster was brought into the inner Milky Way by one of its satellite galaxies, or that its original orbit was perturbed following some dynamical interaction with one or several of them.

The orbit of the Ophiuchus stream is mostly contained in the Galactocentric X--Z plane (the plane containing the Sun, the Galactic centre, and the MW rotation axis), and has an apocentric distance of about 15~kpc. The Sagittarius dSph (Sgr) is a conspicuous candidate for interaction, since it orbits the Galaxy primarily in the same X--Z plane and has a pericentric distance that coincides with that of Ophiuchus' apocentre \citep{gaia18b}.

These may be just coincidences, but they are intriguing enough to warrant further exploration. A full study of all possible GC orbits around Sgr or the Milky Way that may lead to Ophiuchus is beyond the scope of the present work, but we can at least verify the viability of this scenario by assessing whether the presently available data allows for a near passage between Sgr and Ophiuchus in the recent past. This seems like a minimum requirement to argue for a direct connection between Sgr and Ophiuchus.

To investigate this we adopt the kinematics for Sgr from \citet{gaia18b} and sample 1000 sets of phase space coordinates for both Sgr and Ophiuchus, assuming Gaussian uncertainties. We integrate these orbits backwards in \texttt{MWPotential2014} for 500~Myr. We determine at which point Sgr and Ophiuchus come closest to one another and record the time, separation and relative velocity of the encounter. We find that Sgr and Ophiuchus came to within $4.9\pm1.9$~kpc of one another about $96\pm8$~Myr ago, which corresponds to roughly the last apocentric passage of the Ophiuchus stream. At closest approach, Sgr and Ophiuchus had a large relative velocity, of order $279\pm18$~km~s$^{-1}$. Figure~\ref{fig:Sgr} shows the past orbital trace of Sgr (in red) and Ophiuchus (in blue) for 25 example orbits, and highlights the likelihood of a past close encounter.

These findings suggest that while Ophiuchus was likely not originally bound to Sgr due to their large relative velocity, the massive dwarf definitely played a role in shaping the present-day orbit of the stream. It is therefore worthwhile to include Sgr as a gravitating body during future efforts to model Ophiuchus' orbit, especially when the long-term behaviour of the system is under investigation. The scenarios proposed by \citet{price-whelan18} and \citet{hattori16} are both sensitive to the alignment of Ophiuchus' orbit with the galactic bar, suggesting that the interaction with Sgr may require re-assessment of these theories. Including an analytic prescription for Sgr in N-body realizations of Ophiuchus will also highlight any tidal impact Sgr may have had on Ophiuchus during one of their close passages, which could alter the manner in which Ophiuchus disrupts. We plan to pursue this in future work.

We note that our finding that Sgr may have influenced the orbit of Ophiuchus within the time period over which our orbits are integrated should not invalidate our results. The only parameter which we find to be sensitive to the time of disruption is the half-mass radius of the progenitor cluster, which for Ap3 models may be reduced to as low as 10~pc. We therefore propose that if Sgr were to modify the orbit of the progenitor beyond the last 100~Myr, then the effect would likely be to change the inferred size of the progenitor in accordance with the changing tidal field of the resulting orbit.

\section{Summary and Conclusions}
\label{sec:SummaryAndDiscussion}

The Ophiuchus stream is an interesting dynamical puzzle. The observed length of the stream is short, suggesting a recent disruption of the progenitor. On the other hand there is no observed bound core and the stellar population is that of an old metal-poor cluster, suggesting that the stream is much older. One possible resolution to this discrepancy is that the progenitor of this stream is an extremely weakly bound globular cluster, the likes of which are not observed in the Milky Way today.

We have performed a grid search over the possible structural properties of a globular cluster progenitor of the Ophiuchus stream. We evolve these progenitors using N-body simulations to disrupt them along the same orbit as the Ophiuchus stream, and then perform detailed comparisons of the resulting streams to observations.

We find that previously proposed progenitors are too massive to account for the observed properties of the stream. Instead, we find that the width, length, and the number of stars in simulated streams from progenitors with masses of $\sim 2\times10^{3}$~\Msol\, half-mass radii in the range $60$--$100$~pc, which began disrupting about 360~Myr ago yield the best match to observations. There are no known GCs in the Galaxy with these properties, and we speculate that Ophiuchus highlights the presence of yet undiscovered globular clusters in the Milky Way at the faint, low surface brightness end of the GC population.

The Ophiuchus stream may not be unique in this sense. The Phlegethon stream \citep{ibata18} is a stellar stream recently found in the \textit{Gaia} DR2 release, and is thought to have a mass around $1.5\times10^{3}$~\Msol. It may once have been a globular cluster with similar properties to the progenitor of the Ophiuchus stream. The now highly dispersed Phlegethon has an extremely low surface brightness of about $34.6$~mag~arcsec$^{-2}$ in \textit{Gaia} \textit{G}-band. It was only discovered through the use of a dedicated structure-finding algorithm that leverages the full \textit{Gaia} astrometric data set. These types of highly dispersed streams originating from weakly bound globular clusters may be common throughout the Milky Way and remain invisible to us due to their extremely low surface brightnesses.

A cluster as weakly bound as the proposed Ophiuchus progenitor cannot have formed on its current orbit, or anywhere in the inner galaxy for that matter, since it would be susceptible to tidal disruption by the disk and bulge \citep[e.g. see figure 21 in][]{gnedin97}. For the Ophiuchus progenitor to survive to the present day it would have therefore needed to orbit in the outer galaxy, on a low-eccentricity trajectory, for the majority of its $\sim 12$~Gyr life.

If this interpretation is correct, a major question remains: how did Ophiuchus come to orbit where it does today? An interaction with a massive Galactic satellite could provide a possible explanation. We therefore briefly explored the possibility of an interaction between Sgr and the Ophiuchus progenitor, and found that the two passed very close to each other during Ophiuchus last apocentric passage. It is clear that this interaction could have had a substantial effect on Ophiuchus, and that future work will need to consider carefully how the interaction with Sagittarius may have helped to shape the stream properties.

To summarize: we have shown that the progenitor of the Ophiuchus stream likely had a mass of $2\times10^{3}$~\Msol\ or less and half-mass radius in the range $60$-$100$ pc. We find a degeneracy between the size and disruption time of the system, with models with half-mass radii of 60--100~pc which disrupted 360~Myr ago, and denser models with half-mass radii around 10~pc which disrupted 600~Myr ago both providing convincing matches to observations. We obtain our results by analyzing the properties of our simulated streams in a manner consistent with how the real stream was studied. We also perform a basic investigation into the possibility that Ophiuchus has interacted with the Sgr dwarf galaxy in its recent past, and find that the two came to $\sim 5$ kpc from each other about $100$ Myrs ago. It is still unclear what role the bar has played in the evolution of this tidal feature, or how this $\sim12$~Gyr old progenitor came to be on its present orbit. Answers to these questions will require a more detailed modeling of the Galactic potential to include a realistic bar model, as well as a framework to include the influence of Sagittarius on the stream properties.

\section*{Acknowledgements}

The authors would like to thank Jeremy Webb and Eduardo Balbinot for helpful comments which greatly improved the results in this paper. KO received support from VICI grant 016.130.338 of the Netherlands Foundation for Scientific Research (NWO). This project makes use of open source software including \texttt{galpy} \citep{bovy15}, {\sc Gadget-2} \citep{springel05}, {\sc Matplotlib} \citep{hunter07}, and {\sc Astropy} \citep{price-whelan18}. This research has made use of NASA's Astrophysics Data System. This work has made use of data from the European Space Agency (ESA) mission
{\it Gaia} (\url{https://www.cosmos.esa.int/gaia}), processed by the {\it Gaia}
Data Processing and Analysis Consortium (DPAC,
\url{https://www.cosmos.esa.int/web/gaia/dpac/consortium}). Funding for the DPAC
has been provided by national institutions, in particular the institutions
participating in the {\it Gaia} Multilateral Agreement.




\bibliographystyle{mnras}
\bibliography{manuscript}


\appendix

\section{\textit{Gaia} DR2 kinematics}
\label{ap:Gaia}

\begin{figure}
	\includegraphics[width=\linewidth]{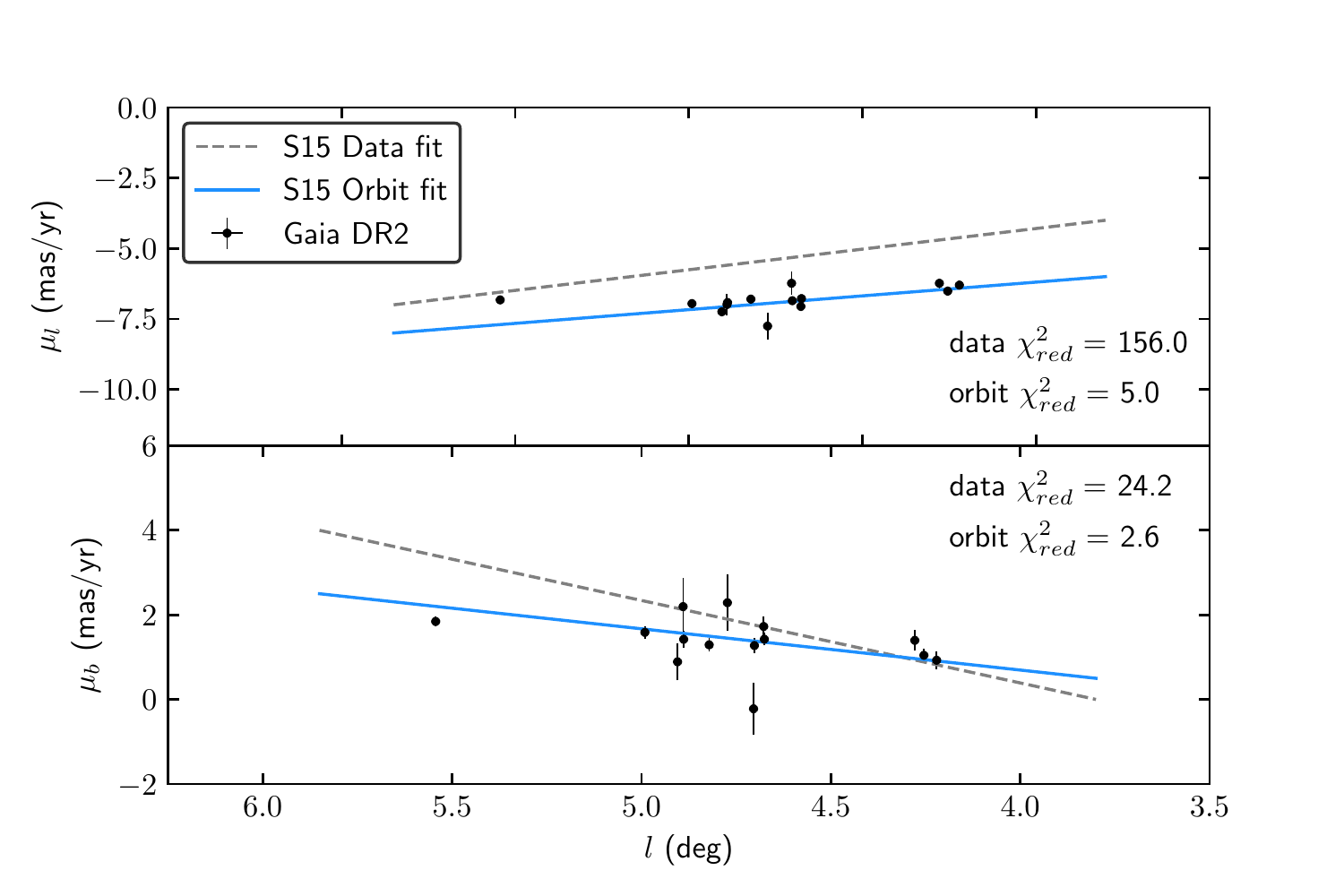}
    \caption{\textit{Gaia} DR2 proper motions as functions of Galactic longitude for stars from \citetalias{sesar15}. The dashed grey and blue lines show the best fit to their data and their best orbital fit, respectively. The \textit{Gaia} DR2 measurements agree very well with the proper motions predicted by the orbit fit in S15.}
    \label{fig:GaiaPM}
  \end{figure}

Measurements for the fourteen stars studied by \citetalias{sesar15} are included in the \textit{Gaia} Data Release 2 \citep{gaia18a}. On-sky positions, parallaxes, proper motions, uncertainties on these quantities, and the proper motion correlation coefficients were obtained from the \textit{Gaia} DR2 archive. Figure~\ref{fig:GaiaPM} shows the proper motions of these objects in Galactic coordinates as a function of Galactic longitude. The straight lines show the best fit to the data from \citetalias{sesar15} (dashed grey) and their best orbital fit (blue).

These new proper motions seem consistent with the \citetalias{sesar15} orbital fit, and in tension with the old proper motion data. As a simple way to confirm this we calculate the reduced Chi-square statistic between the new data and both the orbit and old data linear fits. The results are shown in Figure~\ref{fig:GaiaPM}. The old proper motion data are difficult to reconcile with any reasonable orbit, including one in a non-axisymmetric potential, lending even more weight to the \textit{Gaia} measurements.

\section{Converting particle mass to star counts}
\label{ap:ParticleMassConversion}

In order to facilitate a comparison between simulated and observed streams we must define a correspondence between simulation particles and stream stars. This is achieved by using an isochrone and luminosity function (LF) that match the stellar population of the stream progenitor. The properties of that population -- metallicity, alpha-abundance, age, and the mass-loss parameter -- are all determined by \citetalias{sesar15} and presented in their table~1. We generate an isochrone and LF from the PARSEC v1.2S grid \citep[the same grid used by \citetalias{sesar15};][]{bressan12,chen14,chen15,tang14} assuming these parameters. The LF is calculated using a Chabrier log-normal initial mass function.

To minimize contamination from non-stream stars in their analysis, \citetalias{bernard14} isolate a subset of stars near the main sequence turnoff (MSTO) of the overdensity they detect in colour-magnitude space. Their MSTO selection box lies between PS1 \textit{i}-band magnitudes 20.5 to 18, and spans approximately 0.2 magnitudes in PS1 $g-i$. Using this magnitude information, and the isochrone and LF offset by the mean distance modulus of Ophiuchus, we derive a conversion between particle mass to number of MSTO stars in the following way. The number of MSTO stars in the progenitor system is proportional to the luminosity function, $\Phi$, integrated from 18 to 20.5 magnitudes (about 0.1 \Msol\ \textless\ M \textless\ 0.8 \Msol\ for our isochrone). The mass of the entire globular cluster is the luminosity function times the mass for the corresponding magnitude (inferred from the isochrone), integrated over all magnitudes. The conversion factor can therefore be expressed as the ratio of the above quantities, and we derive this factor to be 0.23 MSTO stars per solar mass.

\begin{equation}
	\frac{N_{\text{MSTO}}}{\rm M_{\odot}} =  \frac{ \int^{18}_{20.5} \Phi(i) {\rm d}i }{ \int_{all} \Phi(i) M(i) {\rm d}i } = 0.23~{\rm M}_{\odot}^{-1}
\end{equation}

In making this conversion we make a number of assumptions. First, that each simulated particle is representative of the entire stellar population. Second, that the mass of the GC is entirely contained in stars that appear in the isochrone, specifically that that there is no dark matter in the globular cluster, that the mass fraction of stars more evolved than the red-giant phase is negligible. We check the resiliency of this conversion to a change in the isochrone and LF grid by performing the same calculation using isochrones and LFs with similar input properties from the Dartmouth Stellar Evolution Database \citep[DSED,][]{dotter08}. The change in the conversion factor is less than 1~per~cent. Our resulting conversion factor agrees heuristically with the results of \citetalias{bernard14}. They find the luminosity of the stream is $\sim1.4\times10^{3}$~L$_{\odot}$, and that there are between 300 and 700 stars in the stream above PS1 \textit{g}-band magnitude of 21 (most of which will be in the MSTO selection box, see \citetalias{bernard14} figure~2c). Assuming our mass-to-light ratio of about 1.45 this equates to about $\sim2\times10^{3}$~\Msol\ and 500 MSTO stars, implying a ratio of 0.25 MSTO stars per M$_{\odot}$.

\subsection{The impact of tidal evolution on the mass function}
\label{ap:MassFunctionFlattening}

A major systematic uncertainty that we must address is how the mass function of a globular cluster evolves in a strong tidal field. It has been well established that the mass function of a globular cluster undergoing tidal stripping is flattened \citep{vesperini97,kruijssen09b,webb15}, decreasing the perceived mass-to-light ratio \citep{anders09,kruijssen09a} and altering its inferred properties \citep[e.g.][]{balbinot18}. Given that the Ophiuchus stream progenitor is thought to be $\sim 11.7$~Gyr old, and its history is poorly understood beyond about a hundred Myr ago, little can be known about the tidal environment in which this cluster has been evolving. If the cluster has been disrupting in place for many Gyr, either being `shepherded' or `fanned' by the bar, it will have been subject to a strong tidal field for the majority of its existence. Conversely, if the system moved onto its present orbit from the outer galaxy, it may have spent most of its life in a weak tidal field. Therefore the exact shape of Ophiuchus mass function is difficult to predict, and is of key importance to inferring the properties of the progenitor.

Here we attempt to estimate the impact of a flattened mass function on the properties of our simulated streams. First, we investigate the effect of flattening on the $N_{\rm MSTO}/{\rm M}_\odot$ conversion factor. We generate a series of isochrone-LF pairs from the DSED database with the same input parameters as presented above, except we now choose a power law initial mass function and vary the power law index between $\alpha=-2.35$ (Salpeter) and $\alpha=0$ (constant number with mass). We compute the conversion factor for each isochrone-LF pair, and find that it varies linearly from 0.12 at $\alpha=-2.35$ to 0.49 at $\alpha=0$. Recall that the value derived above used a Chabrier log-normal initial mass function, which explains why the bottom-heavy Salpeter $\alpha=-2.35$ initial mass function returns such a low conversion factor. Very few globular clusters in the Milky Way have $\alpha>0$, and those few that do have extreme perigalacticon distances of around 1~kpc \citep[e.g. see table 3 of ][ and references therein]{webb15}. We can therefore be confident that the effect of mass function flattening on the conversion from particle mass to $N_{MSTO}$ will be at most an increase by about a factor of 2, from 0.23 to 0.49. We also calculate the V-band mass to light ratio using the same isochrone-LF pairs, and find that it varies from 2.97 for $\alpha=-2.35$ down to 0.61 for $\alpha=0$. Using similar reasoning as above, we can posit that the effect of mass function flattening on the mass to light ratio would be at most a decrease by a factor of about 2.5, from 1.45 to 0.61. We also note that the mass function would not be expected to evolve over the course of our short (< 1~Gyr) simulations, meaning it is not important to distinguish between simulations of different length when considering these effects.

In order to gauge these effects in practice, we replicate our main analysis presented in \S~\ref{sec:ObservingSimulatedStreams} and \ref{sec:Results}, but with the conversion factor set to 0.49 MSTO stars per solar mass, and the cluster mass to light ratio set to 0.61 (as opposed to 0.23 and 1.45 respectively). These parameters were estimated using a mass function power law with index $\alpha=0$, and represent extreme flattening of the mass function. We find that the three matching progenitors now have masses of about $8\times10^{2}$~\Msol, which matches our predictions made above, which were that the inferred mass would be decreased by about a factor of 2. Otherwise the results are nearly identical to those presented in Figure~\ref{fig:ParamVals}, with measured widths, lengths, and velocity dispersions being unchanged, except for the near constant offset in mass. $2\times10^{3}$~\Msol\, is therefore an upper bound on the progenitor mass. For the Ap3 models, when mass function flattening is taken into account, the favoured half-mass radius increases slightly to between 16 and 25~pc, and models with masses lower than $2\times10^{3}$~\Msol\, (but the same half-mass radius) also match within the uncertainties. Given that $2\times10^{3}$~\Msol is already an upper limit for the mass of the progenitor we do not consider these extra models further.


\bsp	
\label{lastpage}
\end{document}